\documentclass[final]{aipproc}

\usepackage{graphicx} 
\usepackage{dcolumn}
\usepackage{bm}
\usepackage{ifpdf}
\layoutstyle{6x9}

\begin{document}

\title{Cosmology in Palatini theories of gravity}



\author{Gonzalo J. Olmo}{
  address={Departamento de F\'{i}sica Te\'{o}rica and IFIC, Centro Mixto Universidad de
Valencia - CSIC. Universidad de Valencia, Burjassot-46100, Valencia, Spain}
}

\begin{abstract}
We discuss recent results on the cosmology of extended theories of gravity formulated in the Palatini approach, i.e., assuming that metric and connection are independent fields. 
In particular, we focus on the attempts to explain the cosmic speedup with $f(R)$ theories and on models that avoid the big bang singularity. The field equations for gravity Lagrangians of the form $f(g_{\mu\nu},{R^\alpha}_{\beta\mu\nu})$ (including torsion) are explicitly derived and discussed. 
\end{abstract}

\maketitle

\section{Introduction}

One of the most elementary and important lessons of Einstein's theory of gravity, general relativity (GR), is that the spatial sections of four dimensional space-time need not be Euclidean. The Minkowskian description is just an approximation  valid on (relatively) local portions of space-time. On larger scales, however, one must consider the deformations induced by the matter on the geometry, which must be dictated by some set of field equations. In this respect, the predictions of GR are in agreement with experiments in scales that range from millimeters to astronomical units, scales in which weak and strong field phenomena can be observed \cite{Will-LR}. The theory is so successful in those regimes and scales that it is generally accepted that it should also work at larger and shorter scales, and at weaker and stronger regimes. The validity of these assumptions, obviously, is not guaranteed {\it a priori} regardless of how beautiful and ellegant the theory might appear to the expert. For this reason, not only must we  keep confronting the predictions of the theory with experiments and/or  observations at new scales, but also we have to demand  theoretical consistency with the other physical interactions and, in particular, in the quantum regime.  \\

Extrapolating the validity of GR at the largest scales forces us to draw a picture of the universe that is not yet supported by other independent observations. For instance, to explain the rotation curves of spiral galaxies, we must accept the existence of vast amounts of unseen matter surrounding those galaxies. Additionally, to explain the luminosity-distance relation of distant type Ia supernovae and some properties of the  distribution of matter and radiation at large scales, we must accept the existence of yet another source of energy with repulsive gravitational properties \cite{reviews}. Together those unseen (or dark) sources of matter and energy are found to make up to $96\%$ of the total energy of the observable universe! This huge discrepancy between the gravitationally estimated amounts of matter and energy and the direct measurements via electromagnetic radiation motivates the search for alternative theories of gravity which can account for the large scale dynamics and structure without the need for dark matter and/or dark energy.\\

On the other hand, the difficulties faced by GR to provide a consistent description of singularities and quantum phenomena at high energies (microscopic or Planck scales) may be an indication that we should go beyond Riemannian structures to successfully quantize the theory and avoid singularities. In fact, the use of the differential geometry of Riemannian spaces in the  formulation of GR soon led to new developments that, in particular, put forward that there is no logical reason to require or impose at a purely kinematical level that the connection should be the Levi-Civita connection of the metric. There is no reason to assume such a strict relation between the metric and affine structures of the theory. 
In this sense, the Palatini formulation of GR was essential in the ADM Hamiltonian formulation of the theory and also played a relevant role in the development of loop quantum gravity, a non-perturbative approach to the quantization of GR. \\

The above arguments support the view that GR may need some kind of extension beyond its usual formulation to address certain observational and theoretical problems. In this work, we review some literature related with the cosmic speedup problem and cosmological singularities  (very early universe) framed within the so-called Palatini formalism, an approach in which  metric and connection are regarded as independent geometrical objects. Though this formulation does not affect the classical dynamics of GR (when torsion is neglected), it does introduce interesting new features as compared to the usual Riemannian approach when extensions of the Einstein-Hilbert Lagrangian are considered.

\section{Dynamics of Palatini theories}

We begin by deriving the field equations of Palatini theories in a very general case and then consider some simplifications to make contact with the literature. For a generic Palatini theory in which the connection appears through the Riemann tensor or contractions of it, the action can be written as follows
\begin{equation}\label{eq:f-action}
S=\frac{1}{2\kappa^2}\int d^4x \sqrt{-g}f(g_{\mu\nu},{R^\alpha}_{\beta\mu\nu})+S_m[g_{\mu\nu},\psi] \ ,
\end{equation}
where $S_m$ is the matter action, $\psi$ represents collectively the matter fields, $\kappa^2$ is a constant with suitable dimensions (if $f=R$, then $\kappa^2=8\pi G$),  and ${R^\alpha}_{\beta\mu\nu}=\partial_\mu\Gamma_{\nu\beta}^\alpha-\partial_\nu\Gamma_{\mu\beta}^\alpha+\Gamma_{\mu\lambda}^\alpha\Gamma_{\nu\beta}^\lambda-\Gamma_{\nu\lambda}^\alpha\Gamma_{\mu\beta}^\lambda$ represents the components of the Riemann tensor, the field strength of the connection $\Gamma^\alpha_{\mu\beta}$. Note that since the connection is determined dynamically, i.e., we assume independence between the metric and affine structures of the theory, we cannot assume any {\it a priori} symmetry in its lower indices. This means that in the variation of the action to obtain the field equations we must bear in mind that $\Gamma^\alpha_{\beta\gamma}\neq  \Gamma^\alpha_{\gamma\beta}$, i.e., we admit the possibility of nonvanishing torsion. It should be noted that in GR energy and momentum are the sources of curvature, while torsion is sourced by the spin of particles
\cite{Kibble-Sciama}. The fact that torsion is usually not consider in introductory courses on gravitation may be rooted in the educational tradition of this subject and the fact that the spin of particles was discovered many years after the original formulation of GR by Einstein. Another reason may be that the effects of torsion are very weak in general, except at very high densities, where the role of torsion becomes dominant and may even avoid the formation of singularities \cite{Hehl-etal,Poplawski}. For these reasons, and to motivate and facilitate the exploration of the effects of torsion in extensions of GR, our derivation of the field equations will be as general as possible (within reasonable limits). \\
 We will assume a symmetric metric tensor $g_{\mu\nu}=g_{\nu\mu}$ and the usual definitions for the Ricci scalar $R\equiv g^{\mu\nu}R_{\mu\nu}$ and Ricci tensor $R_{\mu\nu}\equiv{R^\rho}_{\mu\rho\nu}$. The variation of the action (\ref{eq:f-action}) with respect to the metric and the connection can be expressed as
\begin{eqnarray}\label{eq:var1-f}
\delta S&=&\frac{1}{2\kappa^2}\int d^4x \sqrt{-g}\left[\left(\frac{\partial f}{\partial g^{\mu\nu}} -\frac{f}{2}g_{\mu\nu} \right)\delta g^{\mu\nu} + \frac{\partial f}{\partial {R^\alpha}_{\beta\mu\nu}} \delta {R^\alpha}_{\beta\mu\nu}\right]+\delta S_m \ .
\end{eqnarray}
Straightforward manipulations show that $\delta {R^\alpha}_{\beta\mu\nu}$ can  be written as
\begin{equation}
\delta {R^\alpha}_{\beta\mu\nu}= \nabla_\mu \left(\delta \Gamma^\alpha_{\nu\beta}\right)-\nabla_\nu \left(\delta\Gamma^\alpha_{\mu\beta}\right)+2S^\lambda_{\mu\nu}\delta\Gamma^\alpha_{\lambda\beta} \ ,
\end{equation}
 where $S^\lambda_{\mu\nu}\equiv ( \Gamma^\lambda_{\mu\nu}-\Gamma^\lambda_{\nu\mu})/2$ represents the torsion tensor, the antisymmetric part of the connection. From now on we will use the notation ${P_\alpha}^{\beta\mu\nu}\equiv \frac{\partial f}{\partial {R^\alpha}_{\beta\mu\nu}}$.  In order to put the $\delta {R^\alpha}_{\beta\mu\nu}$ term in (\ref{eq:var1-f}) in suitable form, we need to note that 
\begin{equation}\label{eq:step1}
I_\Gamma=\int d^4x \sqrt{-g} {P_\alpha}^{\beta\mu\nu}\nabla_\mu \delta \Gamma^\alpha_{\nu\beta}=\int d^4x \left[\nabla_\mu(\sqrt{-g}J^\mu)-\delta \Gamma^\alpha_{\nu\beta}\nabla_\mu\left(\sqrt{-g} {P_\alpha}^{\beta\mu\nu}\right)\right] \ ,
\end{equation}
where $J^\mu\equiv {P_\alpha}^{\beta\mu\nu}\delta \Gamma^\alpha_{\nu\beta}$. Since, in general, $\nabla_\mu(\sqrt{-g}J^\mu)=\partial_\mu(\sqrt{-g}J^\mu)+2S^\sigma_{\sigma \mu}\sqrt{-g}J^\mu$, we find that (\ref{eq:step1}) can be written as 
\begin{equation}\label{eq:step2}
I_\Gamma=\int d^4x \left[\partial_\mu(\sqrt{-g}J^\mu)-\delta \Gamma^\alpha_{\nu\beta}\left\{\nabla_\mu\left(\sqrt{-g} {P_\alpha}^{\beta\mu\nu}\right)-2S^\sigma_{\sigma \mu}\sqrt{-g}{P_\alpha}^{\beta\mu\nu}\right\}\right] \ .
\end{equation}
Using this result, (\ref{eq:var1-f}) becomes
\begin{eqnarray}\label{eq:var2-f}
\delta S&=&\frac{1}{2\kappa^2}\int d^4x \left[\sqrt{-g}\left(\frac{\partial f}{\partial g^{\mu\nu}} -\frac{f}{2}g_{\mu\nu} \right)\delta g^{\mu\nu}+\partial_\mu\left(\sqrt{-g}J^\mu\right) \right. \\ 
&+& \left.\left\{-\frac{1}{\sqrt{-g}}\nabla_\mu \left(\sqrt{-g}{P_\alpha}^{\beta[\mu\nu]} \right)+S^\nu_{\sigma\rho}{P_\alpha}^{\beta\sigma\rho}+2S^\sigma_{\sigma\mu}{P_\alpha}^{\beta[\mu\nu]}\right\}2\sqrt{-g}\delta \Gamma^\alpha_{\nu\beta}\right]+\delta S_m \ . \nonumber
\end{eqnarray}
We thus find that the field equations can be written as follows
\begin{eqnarray}\label{eq:gmn}
\kappa^2 T_{\mu\nu}&=&\frac{\partial f}{\partial g^{\mu\nu}} -\frac{f}{2}g_{\mu\nu}  \\
\kappa^2{H_\alpha}^{\nu\beta}&=&-\frac{1}{\sqrt{-g}}\nabla_\mu \left(\sqrt{-g}{P_\alpha}^{\beta[\mu\nu]} \right)+S^\nu_{\sigma\rho}{P_\alpha}^{\beta\sigma\rho}+2S^\sigma_{\sigma\mu}{P_\alpha}^{\beta[\mu\nu]} \ , \label{eq:Gamn}
\end{eqnarray}
where ${P_\alpha}^{\beta[\mu\nu]}=({P_\alpha}^{\beta\mu\nu}-{P_\alpha}^{\beta\nu\mu})/2$, $T_{\mu\nu}=-\frac{2}{\sqrt{-g}}\frac{\delta S_m}{\delta g^{\mu\nu}}$ is the energy-momentum tensor of the matter, and $ {H_\alpha}^{\nu\beta}=-\frac{1}{\sqrt{-g}}\frac{\delta S_m}{\delta \Gamma^\alpha_{\nu\beta}}$ represents the coupling of matter to the connection. For simplicity,   from now on we will assume that  ${H_\alpha}^{\nu\beta}=0$.  Eq. (\ref{eq:Gamn}) can be put in a more convenient form if the connection is decomposed into its symmetric and antisymmetric (torsion) parts, $\Gamma^\alpha_{\mu\nu}=C^\alpha_{\mu\nu}+S^\alpha_{\mu\nu} $, such that $\nabla_\mu A_\nu=\partial_\mu A_\nu-C^\alpha_{\mu\nu} A_\alpha-S^\alpha_{\mu\nu} A_\alpha=\nabla_\mu^C A_\nu-S^\alpha_{\mu\nu} A_\alpha$ and $\nabla_\mu \sqrt{-g}=\nabla_\mu^C \sqrt{-g}-S^\alpha_{\mu\alpha}\sqrt{-g}$. By doing this, (\ref{eq:Gamn}) turns into 
\begin{equation}
\kappa^2{H_\alpha}^{\nu\beta}=-\frac{1}{\sqrt{-g}}\nabla_\mu^C \left(\sqrt{-g}{P_\alpha}^{\beta[\mu\nu]} \right)+S^\lambda_{\mu\alpha}{P_\lambda}^{\beta[\mu\nu]}-S^\beta_{\mu\lambda}{P_\alpha}^{\lambda[\mu\nu]} \ . \label{eq:Gamn2}
\end{equation}

\subsection{Example: f(R,Q) theories}
Eqs. (\ref{eq:gmn}) and (\ref{eq:Gamn2}) can be used to write the field equations for the metric and the connection for specific choices of the Lagrangian $f(g_{\mu\nu},{R^\alpha}_{\beta\mu\nu})$. To make contact with the literature, we now focus on the case $f(R,Q)=f(g^{\mu\nu}R_{\mu\nu},g^{\mu\nu}g^{\alpha\beta}R_{\mu\alpha}R_{\nu\beta})$. 
For this family of Lagrangians, we obtain
\begin{equation}
{P_\alpha}^{\beta\mu\nu}={\delta_\alpha}^\mu M^{\beta\nu}={\delta_\alpha}^\mu\left(f_R g^{\beta\nu}+2f_Q R^{\beta\nu}\right) \ ,
\end{equation}
where $f_X=\partial_X f$. Inserting this expression in (\ref{eq:Gamn2}) and tracing over $\alpha$ and $\nu$, we find that $\nabla_\lambda^C[\sqrt{-g}M^{\beta\lambda}]=(2\sqrt{-g}/3)[S^\sigma_{\lambda\sigma}M^{\beta\lambda}+(3/2)
S^\beta_{\lambda\mu}M^{\lambda\mu}]$. Using this result, the connection equation can be put as follows
\begin{equation}
\frac{1}{\sqrt{-g}}\nabla_\alpha^C\left[\sqrt{-g}M^{\beta\nu}\right]=S^\nu_{\alpha\lambda}M^{\beta\lambda}-S^\nu_{\beta\lambda}M^{\lambda\nu}-S^\lambda_{\alpha\lambda}M^{\beta\nu}+\frac{2}{3}\delta_\alpha^\nu S^\sigma_{\lambda\sigma}M^{\beta\lambda}
\end{equation}
The symmetric and antisymmetric combinations of this equation lead, respectively, to 
\begin{equation}\label{eq:symm}
\frac{1}{\sqrt{-g}}\nabla_\alpha^C\left[\sqrt{-g}M^{(\beta\nu)}\right]=S^\nu_{\alpha\lambda}M^{[\beta\lambda]}-S^\beta_{\alpha\lambda}M^{[\nu\lambda]}-S^\lambda_{\alpha\lambda}M^{(\beta\nu)}+\frac{S^\sigma_{\lambda\sigma}}{3}\left(\delta_\alpha^\nu M^{\beta\lambda}+\delta_\alpha^\beta M^{\nu\lambda}\right)
\end{equation}
and
\begin{equation}\label{eq:asymm}
\frac{1}{\sqrt{-g}}\nabla_\alpha^C\left[\sqrt{-g}M^{[\beta\nu]}\right]=S^\nu_{\alpha\lambda}M^{(\beta\lambda)}-S^\beta_{\alpha\lambda}M^{(\nu\lambda)}-S^\lambda_{\alpha\lambda}M^{[\beta\nu]}+\frac{S^\sigma_{\lambda\sigma}}{3}\left(\delta_\alpha^\nu M^{\beta\lambda}-\delta_\alpha^\beta M^{\nu\lambda}\right)  .
\end{equation}
Important simplifications can be achieved considering the new variables 
\begin{equation}\label{eq:newG}
\tilde{\Gamma}^\lambda_{\mu\nu}=\Gamma^\lambda_{\mu\nu}+\alpha \delta^\lambda_\nu S^\sigma_{\sigma\mu} \ ,
\end{equation}
and taking the parameter $\alpha=2/3$, which implies that $\tilde{S}^\lambda_{\mu\nu}\equiv \tilde{\Gamma}^\lambda_{[\mu\nu]}$ is such that $\tilde{S}^\sigma_{\sigma\nu}=0$. The symmetric and antisymmetric parts of $\tilde{\Gamma}^\lambda_{\mu\nu}$ are related to those of  ${\Gamma}^\lambda_{\mu\nu}$ by
\begin{eqnarray}\label{eq:newC}
\tilde{C}^\lambda_{\mu\nu}&=&C^\lambda_{\mu\nu}+\frac{1}{3}\left(\delta^\lambda_\nu S^\sigma_{\sigma\mu}+\delta^\lambda_\mu S^\sigma_{\sigma\nu}\right) \\
\tilde{S}^\lambda_{\mu\nu}&=&S^\lambda_{\mu\nu}+\frac{1}{3}\left(\delta^\lambda_\nu S^\sigma_{\sigma\mu}-\delta^\lambda_\mu S^\sigma_{\sigma\nu}\right) \label{eq:newS}
\end{eqnarray}
Using these variables, Eqs. (\ref{eq:symm}) and  (\ref{eq:asymm}) boil down to 

\begin{equation}\label{eq:symm2}
\frac{1}{\sqrt{-g}}\nabla_\alpha^{\tilde{C}}\left[\sqrt{-g}M^{(\beta\nu)}\right]=\left[\tilde{S}^\nu_{\alpha\lambda}g^{\beta\kappa}+
\tilde{S}^\beta_{\alpha\lambda}g^{\nu\kappa}\right]g^{\lambda\rho}M_{[\kappa\rho]}
\end{equation}
and
\begin{equation}\label{eq:asymm2}
\frac{1}{\sqrt{-g}}\nabla_\alpha^{\tilde{C}}\left[\sqrt{-g}M^{[\beta\nu]}\right]=\left[\tilde{S}^\nu_{\alpha\lambda}g^{\beta\kappa}-
\tilde{S}^\beta_{\alpha\lambda}g^{\nu\kappa}\right]g^{\lambda\rho}M_{(\kappa\rho)} \ .
\end{equation}
In these equations, $M^{(\beta\nu)}=f_R g^{\beta\nu}+2f_Q R^{(\beta\nu)}(\Gamma)$, and $M^{[\beta\nu]}=2f_Q R^{[\beta\nu]}(\Gamma)$, where $R_{(\beta\nu)}(\Gamma)=R_{(\beta\nu)}(\tilde{\Gamma})$ and $R_{[\beta\nu]}(\Gamma)=R_{[\beta\nu]}(\tilde{\Gamma})-\frac{2}{3}\left(\partial_\beta S^\sigma_{\sigma\nu}-\partial_\nu S^\sigma_{\sigma\beta}\right)$. \\

In the recent literature on Palatini theories, only the torsionless case has been studied in detail. When torsion is considered in $f(R)$ theories, Eqs. (\ref{eq:symm2}) and (\ref{eq:asymm2}) recover the results presented in \cite{Olmo2011review}.  In general, those equations put forward that when the torsion tensor $\tilde{S}^\nu_{\alpha\lambda}$ vanishes, the symmetric and antisymmetric parts of $M^{\beta\nu}$ decouple. The dynamics of these theories, therefore, can be studied in different levels of complexity. The simplest case consists on setting ${S}^\nu_{\alpha\lambda}$ and $R_{[\mu\nu]}$ to zero, which implies that $\Gamma^\sigma_{\sigma\nu}=\partial_\nu \phi$, where $\phi$ is a scalar function that can be determined by solving the equation for $M^{(\mu\nu)}$. In GR, we find that $\phi= \ln \sqrt{-g}$, where $g$ represents de determinant of the metric. In the case of $f(R,Q)$ theories, this result motivates the search for a rank-two tensor $h_{\mu\nu}$ such that $\phi= \ln \sqrt{-h}$. This allows to write the equation  for $M^{\mu\nu}$ in the form $\nabla_\alpha\left(\sqrt{-h}h^{\mu\nu}\right)=0$, which allows to expres  $\Gamma^\sigma_{\mu\nu}$ as the Levi-Civita connection of $h_{\mu\nu}$. The relation between $h_{\mu\nu}$ and $g_{\mu\nu}$ can be explicitly found using (\ref{eq:gmn}) once the matter sources are specified \cite{OSAT}. \\ 
The simplest solution with nonzero torsion corresponds to the case  $\tilde{S}^\nu_{\alpha\lambda}=0$. From (\ref{eq:newS}), we see that the condition $\tilde{S}^\nu_{\alpha\lambda}=0$ implies that the torsion tensor satisfies ${S}^\nu_{\alpha\lambda}=\frac{1}{3}\left({\delta_\alpha}^\nu A_\lambda-{\delta_\lambda}^\nu A_\alpha\right)$, where $A_\alpha\equiv S^\sigma_{\sigma\alpha}$. With this value of the torsion, we can impose  $R_{[\mu\nu]}=0$ if the connection is chosen as  $\Gamma^\sigma_{\sigma\nu}=\frac{2}{3}A_\nu+\partial_\nu \phi$, where $\phi$ is the same scalar function that appears in the torsionless case. It should be noted that the case $\tilde{S}^\nu_{\alpha\lambda}=0$ is the most general solution of the case $R_{[\mu\nu]}=0$.  This follows from (\ref{eq:asymm2}), which can be written as $\left[\tilde{S}^\nu_{\alpha\lambda}g^{\beta\kappa}-
\tilde{S}^\beta_{\alpha\lambda}g^{\nu\kappa}\right]{\Sigma_\kappa}^\lambda =0$ with ${\Sigma_\kappa}^\lambda=f_R{\delta_\kappa}^\lambda+2f_Q {R_\kappa}^\lambda$, and leads to $\tilde{S}^\nu_{\alpha\lambda}g^{\beta\kappa}=\tilde{S}^\beta_{\alpha\lambda}g^{\nu\kappa}$. Contracting this expression with $g_{\beta\kappa}$ leads to $\tilde{S}^\nu_{\alpha\lambda}=0$. When $R_{[\mu\nu]}\neq 0$, then the theory contains new dynamical degrees of freedom \cite{RicciAnti}.

\section{Cosmic speedup in Palatini $f(R)$ theories \label{sec:speedup}}

Observations of the cosmic microwave background (CMB) radiation \cite{CMB}, high redshift supernovae surveys \cite{SNIa}, large scale structure\cite{LSS}, and baryon acoustic oscillations \cite{BAO} suggest that the expansion history of the universe has passed through a number of phases, which consist on an earlier stage of rapidly accelerated expansion (known as inflation) followed by two periods of decelerated expansion dominated by the presence of radiation and dust (matter without pressure), respectively, and a current phase of accelerated expansion that started some five billion years ago following the era of matter domination. The field equations of GR in a Friedmann-Robertson-Walker (FRW) spacetime with line element $ds^2=-dt^2+a^2d\vec{x}^2$ filled with non-interacting perfect fluids of density $\rho_i$ and pressure $P_i$, 
\begin{equation}
\left(\frac{\dot{a}}{a}\right)^2+\frac{K}{a^2}=\frac{\kappa^2}{3}\rho \ , \hspace{0.5cm} \ \frac{\ddot{a}}{a}=-\frac{\kappa^2}{6}(\rho+3P) \ ,
\end{equation}
where $K$ is the spatial curvature, $\rho=\sum_i \rho_i$, and $P=\sum_i P_i$, 
indicate that a phase of positive accelerated expansion can only happen if there exists some matter/energy source that dominates over the others and whose equation of state satisfies $P_X/\rho_X< -1/3$, where $P_X$ and $\rho_X$ represent the pressure and energy density of that source. A natural candidate to explain the current phase of cosmic acceleration is a cosmological constant $\Lambda$, for which $P_\Lambda/\rho_\Lambda=-1$. However, this simple proposal is hard to accept from a theoretical point of view. If $\Lambda$ represents a new fundamental constant of Nature, one could expect new physical phenomena at cosmic scales in analogy with what happened when the Planck constant was discovered. If it is seen as vacuum quantum energy, then it is generally claimed that its observed value is too small to be in agreement with a naive estimation from quantum field theory, though if we apply more rigorous techniques of quantum field renormalization in curved space-times the predicted value turns out to be much smaller \cite{Hollands-Wald} than the observed one. For these and other reasons, there seems to be a widespread desire to explain the current cosmic speedup by means of some dynamical entity rather than by a pure constant of cosmic nature. \\

When the Einstein-Hilbert Lagrangian is extended to a function of the scalar curvature, $f(R)$, eqs. (\ref{eq:gmn}) and (\ref{eq:Gamn}) yield 
\begin{eqnarray}
f_RR_{\mu\nu}-\frac{f}{2}g_{\mu\nu}&=&\kappa^2 T_{\mu\nu} \ , \\
\nabla_\alpha\left[\sqrt{-g}f_Rg^{\mu\nu}\right]&=&0 \ .
\end{eqnarray}
The second equation of above implies that the connection can be solved as the Levi-Civita connection of an auxiliary metric $h_{\mu\nu}=f_R g_{\mu\nu}$, while the trace of the first equation, $R f_R-2f=\kappa^2T$, implies that $R=R(T)$ and $f(R)=f(R(T))$.   The field equations for the metric $g_{\mu\nu}$ can thus be written as 
\begin{eqnarray}\label{eq:Gab-f}
R_{\mu \nu }(g)-\frac{1}{2}g_{\mu \nu }R(g)&=&\frac{\kappa
^2}{f_R}T_{\mu \nu }-\frac{Rf_R-f}{2f_R}g_{\mu \nu
}-\frac{3}{2(f_R)^2}\left[\partial_\mu f_R\partial_\nu
f_R-\frac{1}{2}g_{\mu \nu }(\partial f_R)^2\right]+ \nonumber \\ & &\frac{1}{f_R}\left[\nabla_\mu \nabla_\nu f_R-g_{\mu \nu }\nabla_\lambda \nabla^\lambda
f_R\right] 
\end{eqnarray}
where $R_{\mu \nu }(g)$, $R(g)$, and $\nabla_\mu \nabla_\nu f_R$ are computed in terms of the Levi-Civita connection of the metric $g_{\mu \nu }$, whereas $R$ and $f_R$ must be seen as functions of $T$. The field equations of these theories in vacuum exactly boil down to those of GR with an effective cosmological constant. This turned these theories into a very natural candidate to explain the cosmic speedup. For suitable choices of the function $f(R)$, it could happen that the new gravitationally-induced matter terms that appear on the right hand side of (\ref{eq:Gab-f}) were negligible during earlier phases of the expansion history but became dominant at later times, thus allowing an expansion that closely resembles GR in the past but produces cosmic speedup today. One could thus explain the transition from a matter dominated universe to an asymptotically de Sitter accelerated one with standard sources of matter and radiation but without the theoretical problems posed by a strictly constant $\Lambda$. The most famous $f(R)$ model of this kind investigated in the Palatini approach was borrowed from a proposal of Carroll et al.               \cite{CDTT} in metric formalism, namely, $f(R)=R-\mu^4/R$, where $\rho_\mu\equiv\mu^2/\kappa^2$ represents the energy-density scale at which the effects of the modified dynamics are relevant. Vollick \cite{Vollick-2003} considered this model and showed that after the standard matter-dominated era, the expansion approaches a de Sitter phase exponentially fast. To see this, consider the modified Friedmann equation corresponding to a given $f(R)$ Lagrangian in a universe filled with matter and radiation
\begin{equation}\label{eq:Hubble-iso-f(R)}
H^2=\left(\frac{\dot{a}}{a}\right)^2=\frac{1}{6f_R}\frac{\left[f+\kappa^2(\rho_m+2\rho_r)-\frac{6K f_R}{a^2}\right]}{\left[1+\frac{3}{2}\frac{\kappa^2\rho_m f_{RR}}{f_R(Rf_{RR}-f_R)}\right]^2} \ ,
\end{equation}
where $\rho_m$ represents the energy density of the (pressureless) matter, $\rho_r$ is the energy density of radiation, and $R$ is a function of $\rho_m$ only because $T=-\rho_m$. In the $1/R$ model, one finds 
\begin{equation}\label{eq:R-CDTT}
R=\frac{\kappa^2\rho_m}{2}\left(1+\sqrt{1+12\left(\frac{\rho_\mu}{\rho_m}\right)^2}\right) \ , 
\end{equation}
which recovers $R\approx \kappa^2\rho_m$ when ${\rho_\mu}/{\rho_m}\ll 1$ and tends to the constant value $R_{vac}=\sqrt{3}\mu^2$ when ${\rho_\mu}/{\rho_m}\gg 1$. We thus see that when ${\rho_\mu}/{\rho_m}\ll 1$ then (\ref{eq:Hubble-iso-f(R)}) behaves as
$H^2\approx H^2_{GR}-\kappa^2(\rho_m+4\rho_r/3)(\rho_\mu/\rho_m)^2+\ldots$, which is virtually indistinguishable from GR. However, when the matter energy density, $\rho_m\sim a^{-3}$, drops below the constant value $\rho_\mu$, ${\rho_\mu}/{\rho_m}\gg 1$, then (\ref{eq:Hubble-iso-f(R)}) goes like $H^2\approx \frac{\mu^2}{4\sqrt{3}}+\frac{19}{96}\kappa^2\rho_m+\ldots$, which tends to a constant and implies an asymptotically de Sitter expansion, thus confirming the late time cosmic speedup.\\

The $1/R$ model was soon compared with observations of type Ia supernovae \cite{Meng-Wang}, though such first studies were excessively optimistic about its viability. This optimism may have its origin in earlier studies of Palatini $f(R)$ cosmologies which concluded that these theories were very poorly constrained \cite{BHV-2002}, being $|f_{RR}(0)|<10^{113}$ one of the constraints coming from cosmological data. Besides the $R-\mu^4/R$ theory, which represented a small departure from GR at low matter densities, some authors also explored whether radical departures from the GR dynamics at cosmic scales such as $f(R)=\beta R^n$ or $f(R)=\alpha\ln R$ could be compatible with observations. These models were confronted with the Hubble diagram of type Ia Supernovae, the data on the gas mass fraction in relaxed galaxy clusters \cite{CCF-2004}, and baryon acoustic oscillations \cite{BGS-2006}. Though the fits to the data were good, the statistical analysis did not suggest any improvement with respect to the standard $\Lambda$CDM model. On the other hand, tight constraints on the family of models $R-\alpha R^\beta$ were obtained by studying the cosmic microwave background (CMB) shift parameter and the linear evolution of inhomogeneities \cite{Koivisto-2006} plus the Hubble diagram of type Ia supernovae and baryon oscillations \cite{AEMM-2006}. Besides finding that the $f(R)=R-\mu^4/R$ model was strongly disfavored by the data, it was found that the combined observational data were capable of reducing the allowed parameter space of the exponent $\beta$ to an interval of order $\sim 3 \times 10^{-5}$ around $\beta=0$, with $\alpha$ having a value similar to the cosmological constant. This meant that $R-\alpha R^\beta\approx R-\alpha-\alpha\beta\ln R$ could be restricted to a tiny region around the $\Lambda$CDM model. More stringent constraints on this model were found comparing its predictions with the CMB and matter power spectra \cite{Li:2006ag}, pushing the $\beta$ parameter to the range $\sim 10^{-6}$, thus making this model virtually indistinguishable from $\Lambda$CDM. These conclusions have been reconfirmed by considering updated data \cite{Carvalho:2008am,Santos:2008qp,Pires-2010}  (Union and Union2 supernovae compilations plus other determinations \cite{SJVK2010} of the expansion rate $H(z)$) and strong lensing statistics \cite{YC-2009,Ruggiero:2007jr}. Causality related questions have also been discussed \cite{Santos:2010tw} in relation with this model. A different class of models \cite{Baghram}, with $f(R)=(R^n-R_0^n)^{1/n}$, has recently been confronted with various data samples. The constraints on the parameters, $n=0.98\pm 0.08$, also place this model in the vicinity of the $\Lambda$CDM model.\\
The models considered so far modify the gravitational dynamics at late times, which turns out to be strongly constrained by observations. Modifications at early times should be very weak because of the strong constraints imposed by big bang nucleosynthesis and CMB primary anisotropies. One could thus consider whether modifications at intermediate times could be in agreement with observations. A model proposed in this direction \cite{Li-Chu-2006} takes the form\footnote{A similar model with a $e^{-|R|/(\lambda_2 H_0)^2}/R$ correction was considered in \cite{Ataza2007}. } $f(R)=R+\lambda_1 H_0^2 e^{-|R|/(\lambda_2 H_0)^2}$, where $H_0$ represents the current value of the Hubble parameter, $\lambda_1$ measures the magnitude of the departure from GR, and $\lambda_2$ controls the time at which the correction becomes relevant. Note that at late times this $f(R)$ recovers the $\Lambda$CDM model (which corresponds to the limits $R\to 0$ or $\lambda_2\to \infty$). Though the background evolution of this model is not significantly different from the standard $\Lambda$CDM model for $\lambda_2=500,1000$, which means that it can hardly be constrained by type Ia supernovae data, its effects on the CMB and matter power spectra are dramatic, being $\lambda_2=1000$ safely excluded. The strongest constraints are imposed by the matter power spectrum. This can be understood by looking at the growth
equation for the comoving energy density fluctuations \cite{Koivisto-2006,Koi-Kurki-2006,ULT-2007} $\delta_m$ for large momentum $k$
\begin{equation}
\frac{d^2\delta_m}{dx^2}\approx-\frac{k^2c_s^2}{a^2H^2}\delta_m \ ,
\end{equation}
where $x=\log a(t)$, and $c_s^2=\dot{f}_{R}/(3f_R(2f_R H+\dot{f}_{R}))$ represents the effective sound speed squared. If $c_s^2>0$, the perturbations oscillate instead of growing, whereas for $c_s^2<0$ they become unstable and blow up (this happens for $f(R)=R-\alpha R^\beta$ if $\beta>0$). In the 
$\Lambda$CDM model $c_s^2=0$. The form of the matter power spectrum in the exponential and power-law models, therefore, changes significantly with time developing an intricate oscillatory structure for larger $k$ that clearly conflicts with observations, which allows to strongly constrain the parameter space of the models. The most optimistic constraints restrict the parameter $\lambda_2$ to the region \cite{Li-Chu-2006} $\lambda_2\geq 5\times 10^4$. \\ 
In parallel to the considerations of above, a theoretical consistency check using phase space analysis \cite{FayTT-2006,TUT-2008} was also carried out to determine whether some families of $f(R)$ models could allow for the different phases in the expansion history of the universe suggested by observations. It was shown that radiation, matter, and de Sitter points exist irrespective of the form of the function $f(R)$ provided that the function 
\begin{equation}
C(R)=-3\frac{(Rf_R-2f)Rf_{RR}}{(Rf_{R}-f)(Rf_{RR}-f_R)}
\end{equation}
does not show discontinuous or divergent behaviors. Thus models satisfying the condition $C(R)>-3$ lead to a background evolution comprising the sequence of radiation, matter and de-Sitter epochs. From this it follows that, unlike in metric formalism, theories of the type $f(R)=R-\beta/R^n$ do allow for the sequence of radiation-dominated, matter-dominated, and de Sitter eras if $n>-1$. For theories of the type $f(R)=R+\alpha R^m-\beta/R^n$, one finds that an early inflationary epoch is not followed by a standard radiation-dominated era, which conflicts with the idea that early and late time cosmic acceleration could be unified with this type of models \cite{Sot-2005}. In particular, for $m>2$, the inflationary era is stable and prohibits the end of inflation; if $3/2<m<3$, then inflation ends with a transition to a matter-dominated phase, which is then followed by late time acceleration; for $4/3<m<3/2$, inflation is not possible; and for $0<m<4/3$ one can have the sequence of radiation-dominated, matter-dominated, and late-time de Sitter without early-time inflation. \\ 

\section{Nonsingular bouncing cosmologies \label{sec:QG}}

We have seen that cosmological observations (and local experiments too \cite{LocalExp1,LocalExp2,Olmo-2008a}) strongly constrain the form of the $f(R)$ gravity Lagrangian at low curvatures. Though many $f(R)$ models have the ability to produce late-time cosmic acceleration and fit well the background expansion history, they are not in quantitative agreement with the structure and evolution of cosmic inhomogeneities. On the other hand, it can be shown that the fact that matter is concentrated in discrete structures like atoms causes the modified dynamics to manifest also in laboratory experiments, which confirms earlier suspicions on the viability of such models according to their corresponding Newtonian and post-Newtonian properties. 
This is a very disturbing aspect of the models with infrared corrections, which demands the consideration of a microscopic description of the sources and prevents the use of macroscopic, averaged representations of the matter. A careful analysis of this point put forward the existence of non-perturbative effects induced by the Palatini dynamics in a number of contexts \cite{LocalExp1, Olmo-2008a, Sot_etal_IKPP}. In this sense, it is worth noting that even though the ground state of Hydrogen can be studied using standard perturbative methods, the first and higher excited states do manifest non-perturbative properties \cite{Olmo-2008a}.  
Despite the fact that the modified dynamics is strongly suppressed in regions of high density, non-perturbative effects arise near the zeros of the atomic wavefunctions, where the matter density crosses the characteristic low-density scale of the theory and the gradients of the matter distribution become very important for the dynamics [see eq.(\ref{eq:Gab-f})]. Though this certainly is an undesired property of infrared-corrected models, it could become a very useful tool for models with corrections at high curvatures. Can we construct singularity-free cosmological models that recover GR at low curvatures using the non-perturbative properties of Palatini theories? As we will see, ultraviolet-corrected Palatini models turn out to be very efficient at removing the big bang cosmic singularity in various situations of interest. In this section we will review recent efforts carried out to better understand the properties of Palatini theories in the early universe. \\

\subsection{Non-singular $f(R)$ cosmologies}

Growing interest in the dynamics of the early-universe in Palatini theories has arisen, in part, from the observation that the effective equations of loop quantum cosmology \cite{LQC} (LQC), a Hamiltonian approach to quantum gravity based on the quantization techniques of loop quantum gravity, could be exactly reproduced by a Palatini $f(R)$ Lagrangian \cite{OS-2009}. In LQC, non-perturbative quantum gravity effects lead to the resolution of the big bang singularity by a quantum bounce without introducing any new degrees of freedom. Though fundamentally discrete, the theory admits a continuum description in terms of an effective Hamiltonian that for a homogeneous and isotropic universe filled with a massless scalar field leads to the following modified Friedmann equation
\begin{equation}\label{eq:LQC}
3H^2={8\pi G}\rho\left(1-\frac{\rho}{\rho_{crit}}\right) \ ,
\end{equation}  
where $\rho_{crit}\approx 0.41\rho_{Planck}$. At low densities, $\rho/\rho_{crit}\ll 1$, the background dynamics is the same as in GR, whereas at densities of order $\rho_{crit}$ the non-linear new matter contribution forces the vanishing of $H^2$ and hence a cosmic bounce. This singularity avoidance seems to be a generic feature of loop-quantized universes \cite{Param09}. \\
Palatini $f(R)$ theories share with LQC an interesting property: the modified dynamics that they generate is not the result of the existence of new dynamical degrees of freedom but rather it manifests itself by means of non-linear contributions produced by the matter sources, which contrasts with other approaches to quantum gravity and to modified gravity. This similarity makes it tempting to put into correspondence Eq.(\ref{eq:LQC}) with the corresponding $f(R)$ equation
\begin{equation}
3H^2=\frac{f_R\left(\kappa^2\rho+({R}f_R-f)/2\right)}{\left(f_R-\frac{12\kappa^2\rho f_{RR}}{2 ({R}f_{RR}-f_R)}\right)^2} \ .
\end{equation}
Taking into account the trace equation  for a massless scalar, $R f_R -2f=2\kappa^2\rho$, which implies that $\rho=\rho({R})$, one finds that a Palatini $f(R)$ theory able to reproduce the LQC dynamics (\ref{eq:LQC}) must satisfy the differential equation
\begin{equation}
f_{RR}=-f_R\left(\frac{A f_R -B}{2({R}f_R-3f)A+{R}B}\right)
\end{equation}
where $A=\sqrt{2({R}f_R-2f)(2{R}_c-[Rf_R-2f])}$, $B=2\sqrt{{R}_cf_R(2{R} f_R-3f)}$, and ${R}_c\equiv \kappa^2\rho_c$. If one imposes the boundary condition $\lim_{R\to 0} f_R\to 1$ at low curvatures, and $\ddot{a}_{LQC}=\ddot{a}_{Pal}$ (where $\ddot{a}$ represents the acceleration of the expansion factor) at $\rho=\rho_c$, the solution to this equation is unique.  The solution was found numerically \cite{OS-2009}, though the following function provides a very good approximation from the GR regime to the bouncing region 
\begin{equation}\label{eq:f-guess}
\frac{df}{dR}=- \tanh \left(\frac{5}{103}\ln\left[\left(\frac{R}{12{R}_c}\right)^2\right]\right)
\end{equation} 

Though the function (\ref{eq:f-guess}) implies that the LQC Lagrangian is an infinite series, which is a manifestation of the non-local properties of the quantum geometry, the fact is that one can find non-singular cosmologies of the $f(R)$ type with a finite number of terms. In fact, the quadratic Lagrangian  $f(R)=R+R^2/R_P$ does exhibit non-singular solutions \cite{SS-1990} for certain equations of state \cite{BOSA-2009,MyTalks} depending on the sign of $R_P$. To be precise, if $R_P > 0$ the bounce occurs for sources with $w=P/\rho> 1/3$. If $R_P < 0$, then the bouncing condition is satisfied by $w < 1/3$. This can be easily understood by having a look at the expression for the Hubble function in a universe filled with radiation plus a fluid with generic equation of state $w$ and density $\rho$
\begin{equation}\label{eq:Hubble-iso}
H^2=\frac{1}{6f_R}\frac{\left[f+(1+3w)\kappa^2\rho+2\kappa^2\rho_{rad}-\frac{6K f_R}{a^2}\right]}{\left[1+\frac{3}{2}\Delta_1\right]^2} 
\end{equation}
where ${\Delta}_1=-(1+w)\rho\partial_\rho f_R/f_R=(1+w)(1-3w)\kappa^2\rho  f_{RR}/(f_R(Rf_{RR}-f_R))$. Due to the structure of $\Delta_1$, one can check that $H^2$ vanishes when $f_R\to 0$. A more careful analysis \cite{BO-2010} shows that $f_R\to 0$ is the only possible way to obtain a bounce with a Palatini $f(R)$ theory that recovers GR at low curvatures if $w$ is constant. In the case of $f(R)=R+R^2/R_P$, it is easy to see that $f_R=0$ has a solution if $1+2R_{Bounce}/R_P=0$ is satisfied for $\rho_{Bounce}>0$, where $R_{Bounce}=(1-3w)\kappa^2\rho_{Bounce}$, which leads to the cases mentioned above. \\
Besides avoiding the development of curvature singularities, bouncing cosmologies can solve the horizon problem \cite{Novello-2008}, which makes them interesting as a substitute for inflation. To be regarded as a serious candidate to explain the phenomenology of the early universe, these theories should provide a consistent evolution of perturbations across the bounce, which should also be compatible with the observed nearly scale invariant spectrum of primordial perturbations. Investigations in this direction have found \cite{Koi-2010} that $f(R)$ models that develop a bounce when the condition $f_R=0$ is met turn out to exhibit singular behavior of inhomogeneous perturbations in a flat, dust-filled universe. However, since some terms in the perturbation equations blow up as $f_R\to 0$, their backreaction renders the perturbative system invalid and, therefore, one cannot say if there is a true singularity or not.\\
Further insight on the robustness of the bounce under perturbations  was obtained \cite{BO-2010} studying the properties of $f(R)$ theories in anisotropic spacetimes of Bianchi-I type
\begin{equation}
ds^2=-dt^2+\sum_{i=1}^3 a_i^2(t)(dx^i)^2 \ .
\end{equation} 
If one considers these space-times under the dynamics of Palatini theories with a generic perfect fluid, one can derive a number of useful analytical expressions. In particular, one finds that the expansion $\theta=\sum_i H_i$ and the shear $\sigma^2=\sum_i\left(H_i-\frac{\theta}{3}\right)^2$ (a measure of the degree of anisotropy) are given by 
\begin{equation}\label{eq:Hubble-f(R)}
\frac{\theta^2}{3}\left(1+\frac{3}{2}\Delta_1\right)^2=\frac{f+\kappa^2(\rho+3P)}{2f_R}+\frac{\sigma^2}{2}
\end{equation}
\begin{equation}\label{eq:shear-f(R)}
\sigma^2=\frac{\rho^{\frac{2}{1+w}}}{f_R^2}\frac{(C_{12}^2+C_{23}^2+C_{31}^2)}{3} \ ,
\end{equation}
where the constants $C_{ij}=-C_{ji}$ set the amount and distribution of anisotropy and satisfy the constraint $C_{12}+C_{23}+C_{31}=0$. In the isotropic case, $C_{ij}=0$, one has $\sigma^2=0$ and $\theta^2=9H^2$, with $H^2$ given by Eq.(\ref{eq:Hubble-iso}). Now, 
since homogeneous and isotropic bouncing universes require the condition $f_R=0$ at the bounce, a glance at (\ref{eq:shear-f(R)}) 
indicates that the shear diverges as $\sim 1/f_R^2$. This shows that, regardless of how small the anisotropies are initially, isotropic $f(R)$ bouncing models with a single fluid characterized by a constant equation of state will develop divergences when anisotropies are present. It is worth noting that even though $\sigma^2$ diverges at $f_R=0$, the expansion and its time derivative \cite{BO-2010} are smooth and finite functions at that point if the density and curvature are finite. However, one can check by direct calculation that the Kretschman scalar $R_{\mu\nu\sigma\rho}R^{\mu\nu\sigma\rho}=4(\sum_i(\dot H_i+H_i^2)^2+H_1^2H_2^2+H_1^2H_3^2+H_2^2H_3^2)$ diverges at least as $\sim 1/f_R^4$, which is a clear geometrical pathology and signals the presence of a physical singularity. The problems when $f_R$ vanishes  seem to be generic in anisotropic models of modified theories of gravity \cite{FF-Saa09}. It should be noted, however, that the consideration of several fluids, fluids with varying equation of state, or fluids with anisotropic stresses \cite{Koivisto07}, could affect the dynamics providing new bouncing mechanisms and preventing the extension of this conclusion to such more realistic cases.\\

\subsection{Nonsingular cosmologies beyond $f(R)$ \label{sec:beyond}}

The previous section provides reasons to believe that Palatini $f(R)$ models are not able to produce a fully satisfactory and singularity-free alternative to GR in idealized universes filled with a single perfect fluid with constant equation of state. Though the homogeneous and isotropic case greatly improves the situation with respect to GR, the existence of divergences when anisotropies and inhomogeneities are present spoil the hopes deposited on this kind of Lagrangians. To the light of these results, new Palatini theories were explored \cite{BO-2010} to determine if the introduction of new elements in the gravitational action could avoid the problems that appear in the $f(R)$ models. This led to the study of isotropic and anisotropic cosmologies of some simple generalization of the $f(R)$ family in which the Lagrangian takes the form $f(R,Q)$, with $Q=R_{\mu\nu}R^{\mu\nu}$, and assuming that $R_{[\mu\nu]}=0$. Using the particular Lagrangian 
\begin{equation}\label{eq:f(R,Q)}
f(R,R_{\mu\nu}R^{\mu\nu})=R+a\frac{R^2}{R_P}+\frac{R_{\mu\nu}R^{\mu\nu}}{R_P} \ ,
\end{equation}
where $R_P\sim l_P^{-2}$ is the Planck curvature, it was found that completely regular bouncing solutions exist for both isotropic and anisotropic homogeneous cosmologies filled with a perfect fluid. In particular, one finds that for $a<0$ the interval $0\leq w\leq 1/3$ is always included in the family of bouncing solutions, which contains the dust and radiation cases. For $a\geq 0$, the fluids yielding a non-singular evolution are restricted to $w>\frac{a}{2+3a}$, which implies that the radiation case $w=1/3$ is always nonsingular. For a detailed discussion and classification of the non-singular solutions depending on the value of the parameter $a$ and the equation of state $w$, see \cite{BO-2010}. \\
According to (\ref{eq:gmn}) and (\ref{eq:Gamn2}), the field equations that follow from (\ref{eq:f(R,Q)}) are
\begin{eqnarray}\label{eq:met-var}
f_R R_{\mu\nu}-\frac{f}{2}g_{\mu\nu}+2f_QR_{\mu\alpha}{R^\alpha}_\nu &=& \kappa^2 T_{\mu\nu} \ , \\
\nabla_{\beta}\left[\sqrt{-g}\left(f_R g^{\mu\nu}+2f_Q R^{\mu\nu}\right)\right]&=&0 \label{eq:con-var}
\end{eqnarray}
where $f_R\equiv \partial_R f$ and $f_Q\equiv \partial_Q f$. The connection equation (\ref{eq:con-var}) can be solved in general introducing an auxiliary metric $h_{\alpha\beta}$ such that (\ref{eq:con-var}) takes the form $\nabla_{\beta}\left[\sqrt{-h} h^{\mu\nu}\right]=0$, which implies that $\Gamma^{\rho}_{\mu\lambda}$ can be written as the Levi-Civita connection of $h_{\mu\nu}$. When the matter sources are represented by a perfect fluid, $T_{\mu\nu}=(\rho+P)u_\mu u_\nu+P g_{\mu\nu} $, one can show that $h_{\mu\nu}$ and $h^{\mu\nu}$ are given by \cite{OSAT}
\begin{eqnarray}\label{eq:met-disformal}
h_{\mu\nu}&=&\Omega\left( g_{\mu\nu}-\frac{\Lambda_2}{\Lambda_1-\Lambda_2} u_\mu u_\nu \right)\\
h^{\mu\nu}&=&\frac{1}{\Omega}\left( g^{\mu\nu}+\frac{\Lambda_2}{\Lambda_1} u^\mu u^\nu \right)
\end{eqnarray}
where 
\begin{eqnarray}
\Omega&=&\left[\Lambda_1(\Lambda_1-\Lambda_2)\right]^{1/2} \ , \ \lambda=\sqrt{\kappa^2 P+\frac{f}{2}+\frac{f_R^2}{8f_Q}} \\
\Lambda_1&=& \sqrt{2f_Q}\lambda+\frac{f_R}{2}  \ , \ \Lambda_2= \sqrt{2f_Q}\left[\lambda\pm\sqrt{\lambda^2-\kappa^2(\rho+P)}\right] 
\end{eqnarray}
It is worth noting that (\ref{eq:met-disformal}) implies a disformal relation between the metrics $g_{\mu\nu}$ and $h_{\mu\nu}$. A relation of this form between two metrics naturally arises in Bekenstein's relativistic theory \cite{bekenstein} of MOND and in previous versions of it. In the MOND theory, the vector $u_\mu$ is an independent dynamical vector field and the functions in front of it and in front of $g_{\mu\nu}$ depend on another dynamical scalar field. In the theory described here, on the contrary, the metric tensor is the only dynamical field of the gravitational sector.
Note also that a Palatini-like version of MOND has been recently proposed by Milgrom \cite{Milgrom:2009ee}.\\
In terms of $h_{\mu\nu}$ and the above definitions,  (\ref{eq:met-var}) becomes
\begin{equation}\label{eq:Rmn-h}
R_{\mu\nu}(h)=\frac{1}{\Lambda_1}\left[\frac{\left(f+2\kappa^2P\right)}{2\Omega}h_{\mu\nu}+\frac{\Lambda_1\kappa^2(\rho+P)}{\Lambda_1-\Lambda_2}u_{\mu}u_{\nu}\right] \ .
\end{equation}
In this expression, the functions $f, \Lambda_1$, and $\Lambda_2$ are functions of the density $\rho$ and pressure $P$. In particular, for our quadratic model one finds that $R=\kappa^2(\rho-3P)$, like in GR, and $Q=Q(\rho,P)$ is given by 
\begin{equation}\label{eq:Q}
\frac{Q}{2R_P}=-\left(\kappa^2P+\frac{\tilde f}{2}+\frac{R_P}{8}\tilde f_R^2\right)+\frac{R_P}{32}\left[3\left(\frac{ R}{R_P}+\tilde f_R\right)-\sqrt{\left(\frac{R}{R_P}+\tilde f_R\right)^2-\frac{ 4 \kappa^2(\rho+P)}{R_P} }\right]^2 \ ,
\end{equation}
where $\tilde f=R+aR^2/R_P$, and the minus sign in front of the square root has been chosen to recover the correct limit at low curvatures. In a universe filled with radiation, for which $R=0$, the function $Q$ boils down to \cite{BO-2010}  
\begin{equation}
Q= \frac{3R_P^2}{8}\left[1-\frac{8\kappa^2\rho}{3R_P}-\sqrt{1-\frac{16\kappa^2\rho}{3R_P}}\right] \label{eq:Q-rad} \ .
\end{equation}
This expression recovers the GR value at low curvatures, $Q\approx 4(\kappa^2\rho)^2/3+32(\kappa^2\rho)^3/9R_P+\ldots$ but reaches a maximum $Q_{max}=3R_P^2/16$ at $\kappa^2\rho_{max}=3R_P/16$, where the squared root of (\ref{eq:Q}) vanishes. At $\rho_{max}$ the shear also takes its maximum allowed value, namely, $\sigma^2_{max}=\sqrt{3/16}R_P^{3/2}(C_{12}^2+C_{23}^2+C_{31}^2)$, which is always finite, and the expansion vanishes producing a cosmic bounce regardless of the amount of anisotropy. The model (\ref{eq:f(R,Q)}), therefore, avoids the well-known problems of anisotropic universes in GR \cite{ekpyrotic}, where anisotropies grow faster than the energy density during the contraction phase leading to a singularity that can only be avoided by sources with $w>1$. \\

The evolution of inhomogeneities in the quadratic model discussed here was considered in  \cite{LBM07}, though the approximations used there to solve for the connection equation did not allow to see the existence of bouncing solutions. For this reason, in this case one cannot make any statement regarding the evolution of inhomogeneities across the bounce. The cosmology of $f(R)$ and $f(R_{\mu\nu}R^{\mu\nu})$ theories was also considered in some detail in \cite{ABF04}. The possibility of having a standard cosmological evolution in $f(R,Q)$ models with a large cosmological constant has been considered recently\cite{Bauer:2010bu}. \\
Before concluding, we note that the $f(R,Q)$ theories discussed here are able to reproduce other aspects of the expected phenomenology of quantum gravity at the Planck scale \cite{Olmo-JCAP}. In particular, without imposing any a priori phenomenological structure, the quadratic Palatini model (\ref{eq:f(R,Q)}) predicts an energy-density dependence of the metric components that closely matches the structure conjectured in models of Doubly (or Deformed) Special Relativity \cite{DSR} and Rainbow Gravity \cite{RG}. This confirms that Palatini theories represent a new and powerful framework to address different aspects of quantum gravity phenomenology.

\begin{theacknowledgments}
Work supported by the Spanish grant FIS2008-06078-C03-02 and the Consolider Programme CPAN (CSD2007-00042).
\end{theacknowledgments}

\bibliographystyle{aipproc}

\end{document}